\documentclass{aa}  

\usepackage{graphicx}
\usepackage{ulem}
\usepackage{txfonts}
\usepackage{xcolor}
\usepackage{hyperref}

\begin{document}

   \title{Luminosity dependence of the multiple cyclotron lines in \\ 4U 0115$+$63}

   \subtitle{}

   \author{Kinjal Roy$^1$\thanks{kinjal@rrimail.rri.res.in},
          Hemanth Manikantan$^{1,2}$,
          \and
          Biswajit Paul$^1$
          }

   \institute{$^1$Raman Research Institute, C. V. Raman Avenue, Sadashivanagar, Bengaluru - 560 080, India.\\
   $^2$INAF - Istituto di Astrofisica e Planetologia Spaziali, Via del Fosso del Cavaliere, 100, 00133 Roma, Italy. \\
             }

   \date{}

 
  \abstract
   {The Be X-ray binary 4U 0115$+$63 underwent a giant outburst in 2023 with the X-ray luminosity of the source reaching 10$^{38}$ erg s$^{-1}$. During the outburst, two target of opportunity observations were made with \textit{NuSTAR}.}
   {The main goal of this work is to model the multiple cyclotron scattering features (CRSFs) present in 4U 0115$+$63 and study their dependence on the luminosity of the source.}
   {The 3$-$79 keV broadband X-ray coverage of \textit{NuSTAR} allowed us to properly model the continuum and investigate the nature of the multiple cyclotron resonance scattering features present in the source spectrum. We used the epoch-folding technique to find the pulsation from the source and also studied the changes in the cyclotron line energy with an order of magnitude variation in the source luminosity.}
   {We detected five cyclotron lines during the 2023 outburst near 12, 16, 24, 34, and 47 keV. The $\sim$16 keV cyclotron line cannot be harmonically related to the other cyclotron lines at $\sim$12 keV and $\sim$24 keV. This indicates the presence of two fundamental lines in the spectrum of 4U 0115$+$63 at 12 keV and 16 keV.}
   {With the inclusion of the two latest \textit{NuSTAR} observations, we have expanded the data set of the CRSF line center to encompass a broad range of luminosity. This enables us to comprehensively investigate the relationship between the centroid energy of the cyclotron lines and luminosity. The CRSF line center shows no anticorrelation with luminosity, unlike previously reported. Instead, a weak positive correlation is found in four out of the five detected cyclotron lines of 4U 0115$+$63. The luminosity variation of the two fundamental CRSFs could be well explained with the prediction from the collisionless shock model. A tentative negative correlation was observed in the fundamental CRSF at 16 keV and its harmonics beyond a "critical luminosity" of 10$^{38}$ erg/s. This behavior was not present for the 12 keV fundamental CRSF and its harmonic at 24 keV.}


   \titlerunning{Luminosity dependence of cyclotron lines in 4U 0115$+$63}
   \authorrunning{Roy et. al.} 
   \maketitle

%

\section{Introduction}

4U 0115$+$63 is a transient high mass X-ray binary (HMXB) pulsar that was discovered during the sky survey conducted with the \textit{Uhuru} mission~\citep{Uhuru_catalogue_Giacconi_1972}. The source was subsequently discovered to be a Be X-ray binary system with the main sequence star identified as a B0.2Ve spectral-type star~\citep{Johns_1978}. The orbital period of 4U 0115$+$63 is $\sim$ 24.3 days~\citep{Rappaport_1978}. The neutron star (NS) at the heart of the Be X-ray binary system has a spin period of $\sim$ 3.6 s~\citep{Cominsky_1978}. The distance to the binary system has been estimated to be $\sim$ 7 kpc~\citep{0115p63_distance_Negueruela_2001}. 4U 0115$+$63 has gone through many episodes of type II outbursts, most recently in 2011~\citep{Suzaku_2008_outburst_Iyer_2015}, 2015~\citep{NuSTAR_2015_outburst_Liu_2020} and 2023~\citep{4U0115p63_2023_outburst_ATel_MAXI}. The latest outburst occurred during March 2023 with the X-ray luminosity reaching $10^{38}$ erg s$^{-1}$~\citep{4U0115p63_2023_outburst_ATel_MAXI}.

In highly magnetized (B $\ge$ $ 10^{12}$ G ) neutron stars in binary systems, accreted matter gets channeled along the magnetic field lines to the poles of the NS. A hot spot is produced on the accretion mound near the magnetic poles of the NS giving rise to strong X-ray emissions~\citep{Becker_2007}. The presence of strong magnetic fields at the polar regions lead to quantization of electron energy levels ($E_n$ in keV) in accordance to the Landau levels~\citep{Meszaros_1992_book}

\begin{equation}
E_n = \frac{n}{(1+z)} \frac{eB\hbar}{m_e c} \sim \frac{11.6 \times n \times B_{12}}{1+z}
\label{E_cyc_vs_B_12}
,\end{equation}

where $m_e$ is the mass of an electron, $B_{12}$ is the magnetic field strength in units of $10^{12}$ G, $n$=1,2,3,4,.. denotes the different levels, and z is the gravitational redshift due to the NS mass. The cyclotron resonant scattering feature (CRSF) is the absorption feature in the spectra of pulsars due to interactions between photons and electrons in quantized energy levels present in the line-forming region of the NS. This spectral feature facilitates the direct measurement of the magnetic field strength of the NS.  While some NS spectra do not contain any CRSFs, the sources that do contain CRSFs -- the exact region responsible for the production of CRSFs -- are still under debate~\citep{CRSF_review_Staubert_2019}. Also, we do not have a complete picture when it comes to the variation of the cyclotron line with X-ray luminosity. Cyclotron line energy has been shown to be correlated with luminosity~\citep{CRSF_review_Staubert_2019}. One such model used to describe the variation in the cyclotron line energy with luminosity is the shock height model~\citep{CRSF_Becker_2012} where the height of the CRSF formation region varies with luminosity. The cyclotron line energy and source luminosity have a positive correlation when the NS is in Coulombic shock regime. However, beyond a certain 'critical' luminosity, the parameters show negative correlation as the NS enters the radiative shock regime. Such a transition from a positive to negative correlation beyond a critical luminosity has been observed in A 0535$+$26~\citep{A0535p26_Kong_2021} and V 0332$+$53~\citep{V0332p53_CRSFvslumin_Lutovinov_2015, V0332p53_Doroshenko_2017}. A\ transition over critical luminosity is sometimes associated with a change in the pulse profile~\citep{V0332p53_CRSFvslumin_Lutovinov_2015, Critical_lumin_WilsonHodge_2018}. An alternate behavior between the CRSF energy and luminosity has been observed in GX 304$-$1~\citep{GX304m1_CRSF_Rothschild_2017} and Cep X$-$4~\citep{CepX4_CRSF_Vybornov_2017} where the CRSF energy plateaus at higher values of luminosity.

Multiple cyclotron lines have been observed in 4U 0115$+$63 near 11, 23, 33, 41, and 53 keV~\citep{Wheaton_1979, White_1983, Heindl_1999, Santangelo_1999, Ferrigno_2009}. The harmonics of a fundamental CRSF line has been seen in multiple X-ray pulsars such as Vela X-1~\citep{Kendziorra_VelaX1_CRSF_1992, Kretschmar_VelaX1_CRSF_1996, Vela_X1_2CRSF_Kreykenbohm_2002, Chandreyee_Vela_X1_CRSF_2013}, Centaurus X-3~\citep{Nagase_CenX3_CRSF_1992, Santangelo_CenX3_CRSF_1998, Cen_X3_2CRSF_Yang_2023}, Cep X-4~\citep{Mihara_CepX4_CRSF_1991, McBride_CepX4_CRSF_RXTE_2007,  CepX4_CRSF_Vybornov_2017}, and 4U 1907+09~\citep{Cusumano_4U1907p09_CRSF_1998, Makishima_4U1907p09_CRSF_1999, 4U1907p09_2CRSF_2023}. From the analysis of \textrm{GINGA} spectra of 4U 0115$+$63 at different luminosities during the 1991 outburst, it was reported that the CRSF line energy increased from 12 keV to 17 keV with a decrease in X-ray luminosity~\citep{Mihara_1998}. An anticorrelation between centroid energy of the $\sim$11 keV CRSF line with luminosity was reported with \textrm{RXTE} data during the 1999 outburst~\citep{Nakajima_2006}. However, a subsequent analysis of data from the 2008 outburst of 4U 0115$+$63 has put the validity of this claim~into question \citep{Muller_2013, Boldin_2013}. A paradigm shift occurred regarding the understanding of CRSFs in 4U 0115$+$63 when the continuum model was modified by ~\citet{Suzaku_2008_outburst_Iyer_2015} such that two lines near 12 and 15 keV were detected simultaneously along with their higher harmonics from \textrm{RXTE} and \textit{Suzaku} data. Subsequent observations of 4U 0115$+$63 using \textit{NuSTAR} during the 2015 outburst~\citep{NuSTAR_2015_outburst_Liu_2020} firmly established the presence of two fundamental CRSFs near 12 and 16 keV.

\textit{NuSTAR} made four Target of Opportunity (ToO) observations of the source during the latest type II outburst in 2023. One observation was made  at the peak of the outburst, another during the decay phase of the outburst, while two other observations were made when the source went back to quiescence. In this paper, we report the results from the first two observations taken at the peak and decay phase of the outburst. We report on the evolution of CRSFs with luminosity of the pulsar, using the updated spectral model of the source with a simultaneous fitting of the 12 and 16 keV CRSF.

\section{Observation and data reduction}

\begin{figure}
    \centering
    \includegraphics[width=0.45\textwidth]{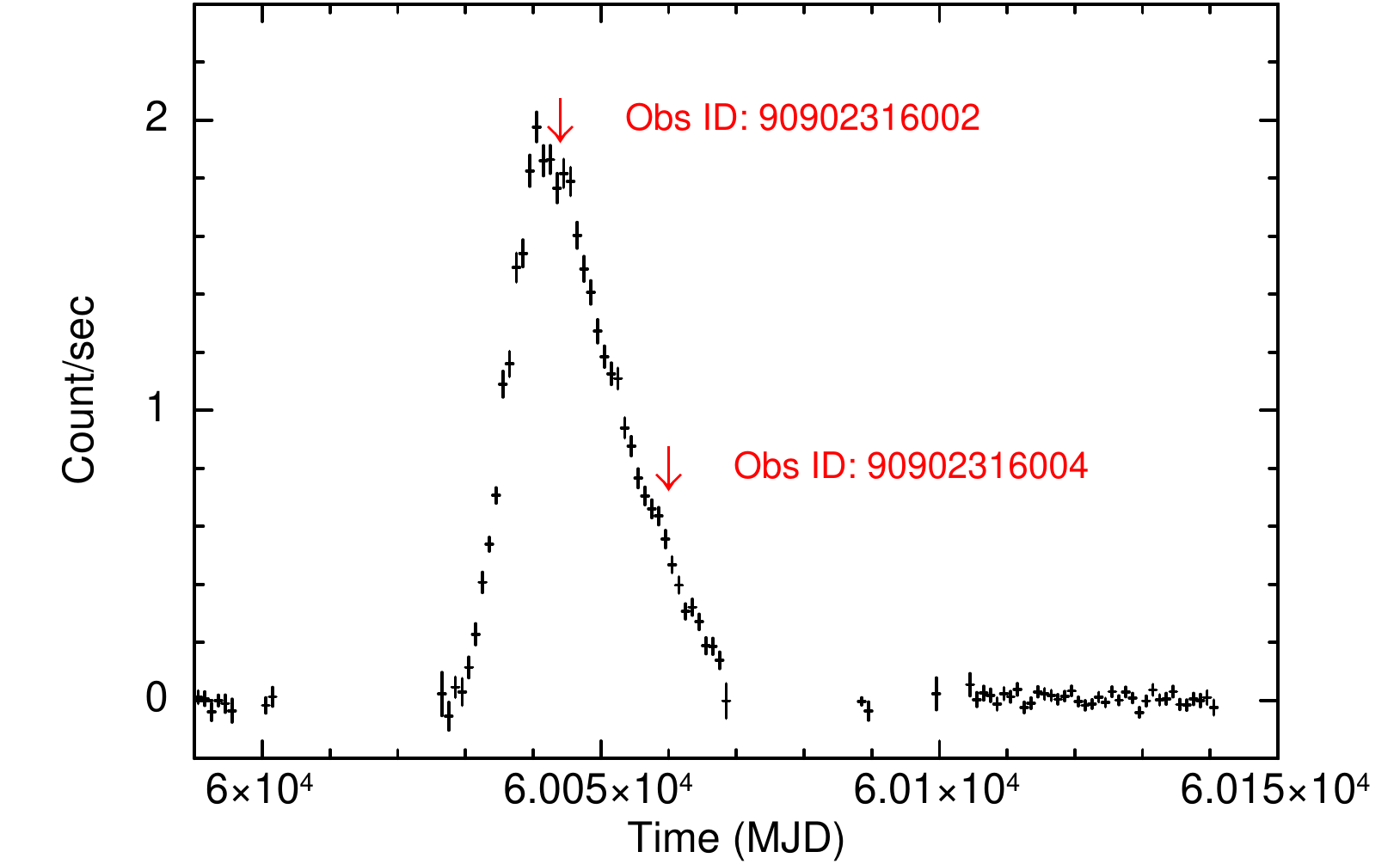}
    \caption{2023 outburst of 4U 0115$+$63 as seen by \textrm{MAXI}/GSC. In black is the 2-20 keV \textrm{MAXI}/GSC count rate of the source binned to one day intervals. The two red arrows point to the two \textit{NuSTAR} observations of the source analyzed in this work.
    \label{fig:Outburst_NuSTAR_obs}}
\end{figure}

The first observation by \textit{NuSTAR} for the 2023 outburst of 4U 0115$+$63 occurred on 9 April 2023 (MJD 60043) (Obs ID: 90902316002) and the second one on 26 April 2023 (MJD 60060) (Obs ID: 90902316004). Hereafter Obs ID: 90902316002 and 90902316004 are abbreviated as ObsID 02 and ObsID 04. The times of the two ToO observations are given in Fig.~\ref{fig:Outburst_NuSTAR_obs} along with the 2$-$20 keV \textrm{MAXI}/GSC light curve of the source. The first observation was taken at the peak, while the second observation was performed during the decay phase.

The \textit{NuSTAR} data were reduced using \texttt{HEASOFT} version \texttt{6.30.1} along with the latest calibration files available via \texttt{CALDB}. The \texttt{nupipeline} version \texttt{0.4.9} was used to produce the clean event files, while the \texttt{nuproduct} command was used to extract the light curves and spectrum corresponding to the source and background regions. A circle with a 150 arcsec radius around the source position was used for extracting the source level 2 data; similarly, a circle with the same radius away from the source region was used to extract the background data. Solar system barycenter correction was performed using \texttt{barycorr} version \texttt{2.16}. All the spectral files were optimally binned based on~\citet{Optimal_binning_Kaastra_Bleeker_2016}.  The light curve for the 2023 outburst was obtained from \texttt{MAXI}~\footnote{\url{http://maxi.riken.jp/mxondem/}} \citep{maxi_main_paper} in the 2-20 keV energy range binned to intervals of one day.

\section{Spectral and timing analysis}

\subsection{Timing analysis}

\begin{figure*}
    \centering
    \includegraphics[width=0.95\textwidth]{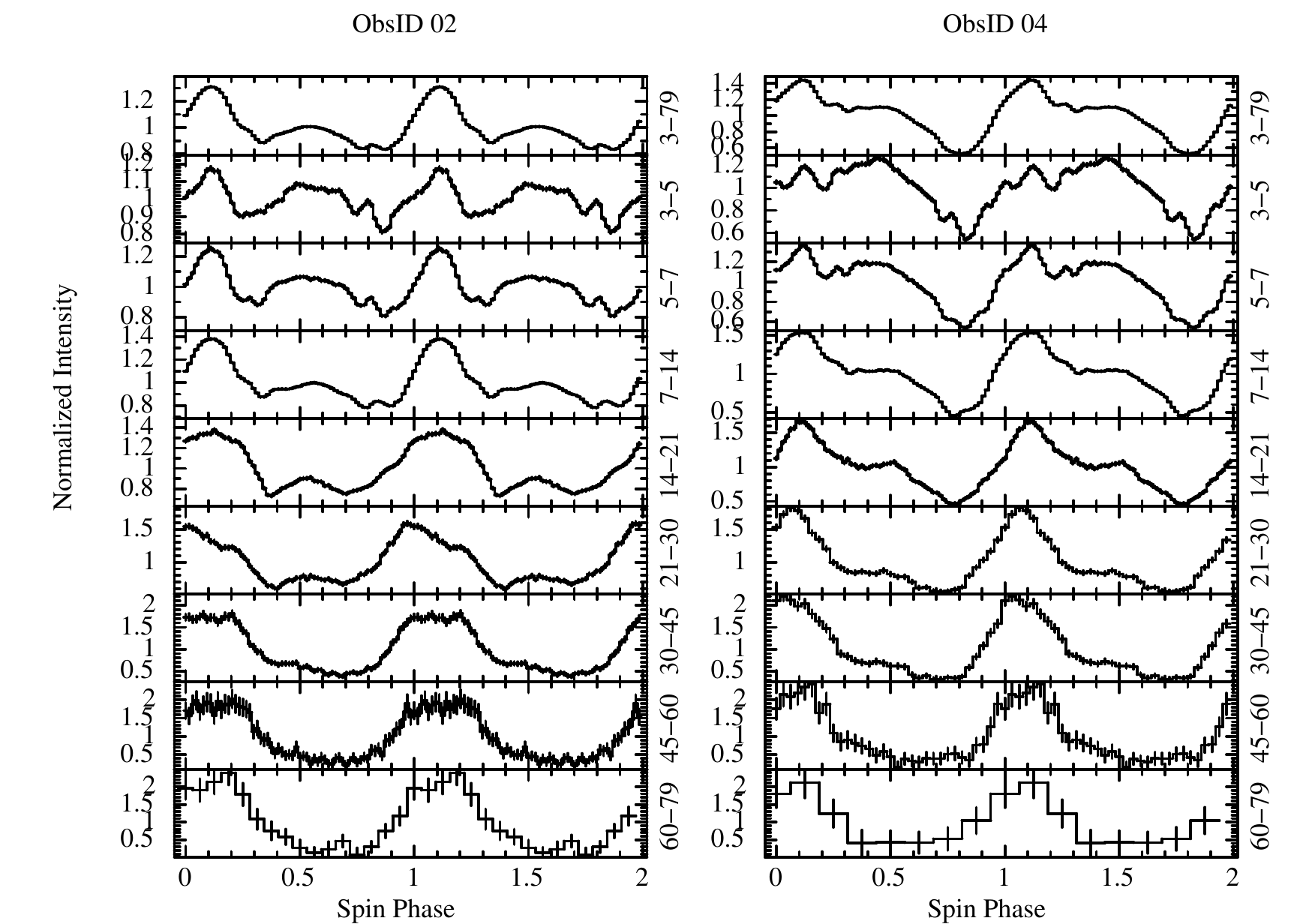}
    \caption{Folded profile of 4U 0115$+$63 from the two observations during the 2023 outburst normalized by dividing the average count rate from each profile. From the top down, we plot the overall (3$-$79 keV) and energy resolved (3$-$5, 5$-$7, 7$-$14, 14$-$21, 21$-$30, 30$-$45, 45$-$60, and 60$-$79 keV) pulse profiles for ObsID 02 \textbf{(Left Panel}) and ObsID 04 \textbf{(Right Panel)}.
    \label{fig:NuSTAR_energy_resolved_pulse_profile}}
\end{figure*}

The broadband (3$-$79 keV) FPMA and FPMB averaged count rate of the source changed from 444 cts/s for ObsID 02 to 114 cts/s in the second observation, which is a decrease by a factor of $\sim$ 4. The spin period of the source was found to be 3.61456(1) s for the first observation ObsID 02 using the epoch-folding tool \texttt{efsearch}. While for the second observation, ObsID 04, the spin period was reduced to 3.61309(1) s. The 3$-$79 keV pulse profiles of the two observations are shown in the top panel of Fig.~\ref{fig:NuSTAR_energy_resolved_pulse_profile} folded with an arbitrary epoch to align the peaks. The pulse profiles were found to be broadly double peaked for both the observations with the first peak near phase $\sim$0.1 and the second peak near phase $\sim$ 0.55. The dip between the two peaks is more prominent in ObsID 02 and there is also a hint of a smaller third peak in the same observation near the pulse phase $\sim$ 0.8. 

We extracted the source light curve of 4U 0115$+$63 for multiple energy ranges to study the variation in the pulse profile with energy. The folded profiles in the energy ranges of 3$-$5, 5$-$7, 7$-$14, 14$-$21, 21$-$30, 30$-$45, 45$-$60, and 60$-$79 keV for ObsID 02 (Left Panel) and ObsID 04 (right panel) are given in  Fig.~\ref{fig:NuSTAR_energy_resolved_pulse_profile}. The pulse profile of 4U 0115$+$63 shows significant evolution with energy, as well as luminosity. The profiles show a significant change between the two observations with the pulse shape evolving from multiple peaks at lower energies to a single peak at higher energies. The pulse fraction shows an increasing trend with energy for both observations.

\subsection{Spectral analysis}

The spectra from FPMA and FPMB for 4U 0115$+$63 were fitted together allowing for a relative normalization \citep{FPM_crosscal_Madsen_2015}, while all other parameters were tied between the two spectra. All of the spectral analyses were carried out using \texttt{XSPEC} version \texttt{12.12.1} \citep{XSPEC} with the absorption by the interstellar medium modeled using the Tuebingen-Boulder absorption model \texttt{TBabs} and elemental abundances being taken from~\citet{Wilm_abund} and the photoelectric cross sections from~\citet{Vern_cs}.

The \textit{NuSTAR} broadband continuum spectrum of ObsID 02 was fitted with the \texttt{compTT} spectral model describing the Comptonization of soft photons in a hot plasma~\citep{comptt_Titarchuk_1994}. A soft excess was present at lower energies which was modeled by a blackbody component (\texttt{bbodyrad}) with a temperature of $\sim$0.6 keV. The fluorescent iron K$\alpha$ line present in the spectrum was modeled using a \texttt{Gaussian} emission line at $\sim$6.4 keV. The continuum model was multiplied by a \texttt{tbabs} spectral model to take into account the absorption of soft X-rays by interstellar matter. However, there were prominent absorption features still present in the residuals of the best fit model. We used the common absorption model \texttt{gabs} to model the different absorption features present in the spectrum. The final spectral model used was \texttt{tbabs} $\times$ ( \texttt{compTT} + \texttt{bbodyrad} + \texttt{Gaussian} ) $\times$ multiple \texttt{gabs}.

\begin{figure}
    \centering
    \includegraphics[width=0.45\textwidth]{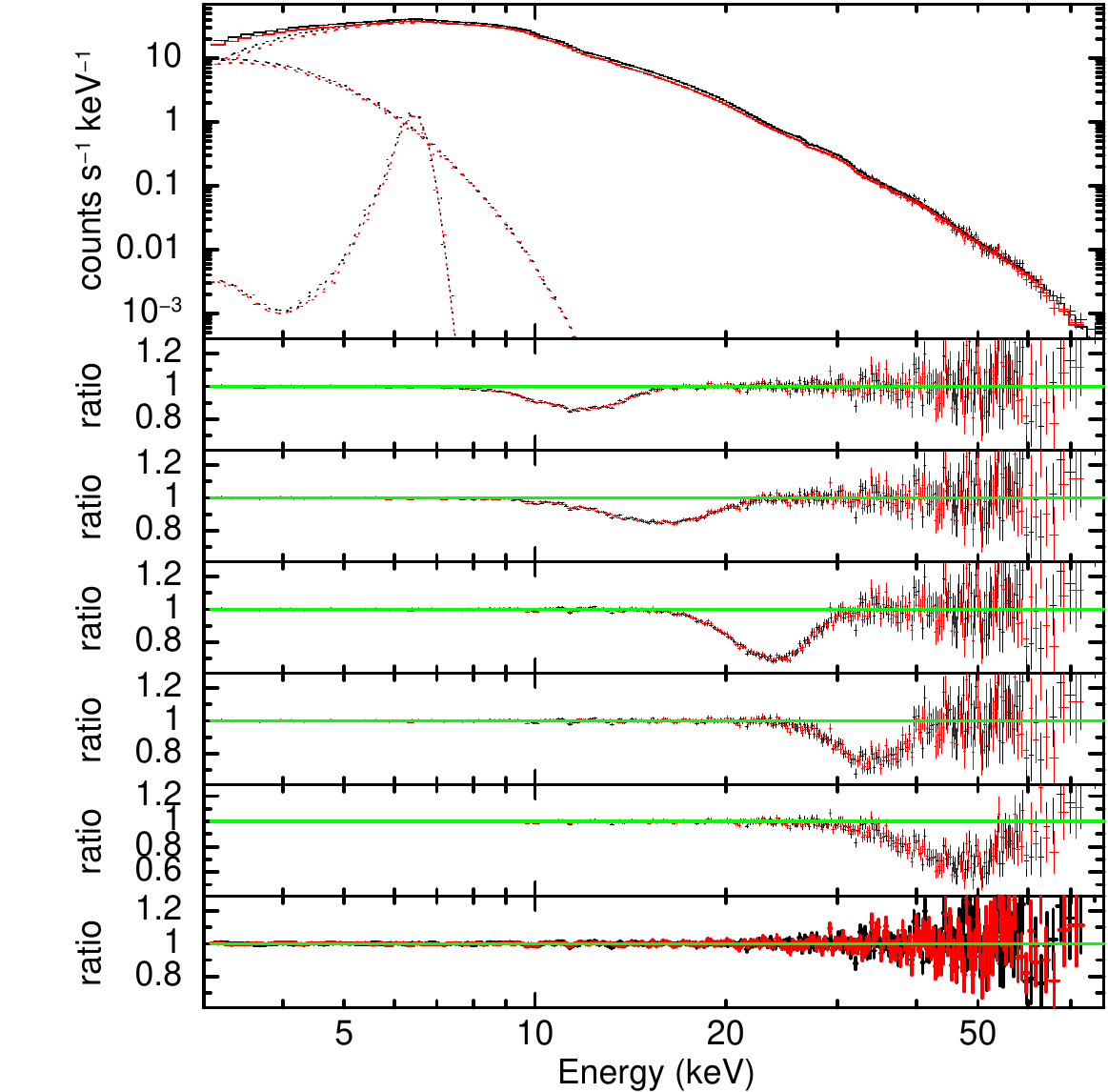}
    \caption{Spectrum and best fit spectral model for ObsID 02. The top panel consists of the best fit models along with the different components. Below this, the next five panels correspond to the residue from each of the five CRSFs at $\sim$12 keV, $\sim$16 keV, $\sim$24 keV, $\sim$34 keV, and $\sim$47 keV that were removed individually from the best fit model by setting the corresponding depth to zero. The bottom panel corresponds to the best fit model residue. The black points correspond to FPMA, while red points are for FPMB.
    \label{fig:Spectrum_2023_ID_2}}
\end{figure}

The spectrum of ObsID 02 (Fig.~\ref{fig:Spectrum_2023_ID_2}) showed two broad features around 15 and 25 keV and two narrower features around 34 and 47 keV. The broad feature around 15 keV could not be fitted with one single broad absorption feature. Therefore, we fitted two Gaussian absorption lines around this energy range, which reduced the $\chi^2$ from 775 for 424 degrees of freedom to 542 for 421 degrees of freedom. The five distinct absorption features (modeled as \texttt{gabs})~\footnote{\url{https://heasarc.gsfc.nasa.gov/}}~\footnote{\url{https://heasarc.gsfc.nasa.gov/docs/xanadu/xspec/index.html}} could be identified as five CRSF lines around $\sim$ 12, 16, 24, 34, and 47 keV.

The two CRSFs present at $\sim$12 keV and $\sim$16 keV cannot be harmonics from the same fundamental line. Therefore, the two CRSFs at around 12 and 16 keV must both be fundamental lines. The CRSF at $\sim$24 keV is the first harmonic of the $\sim$12 keV line and the line at $\sim$34 keV is the combination of the second harmonic to the 12 keV fundamental and the first harmonic to the 16 keV fundamental, while the $\sim$47 keV line is the combination of the third harmonic to the 12 keV fundamental and the second harmonic to the 16 keV fundamental. The best fit model gave a reduced $\chi^2$ of 1.29 for 421 degrees of freedom. The luminosity of the source was calculated to be $12.08^{+0.02}_{-0.06}$ $\times$ $10^{37}$ erg s$^{-1}$ in the 3$-$50 keV energy range, assuming a distance of 7 kpc~\citep{0115p63_distance_Negueruela_2001}.

\begin{figure}
    \centering
    \includegraphics[width=0.45\textwidth]{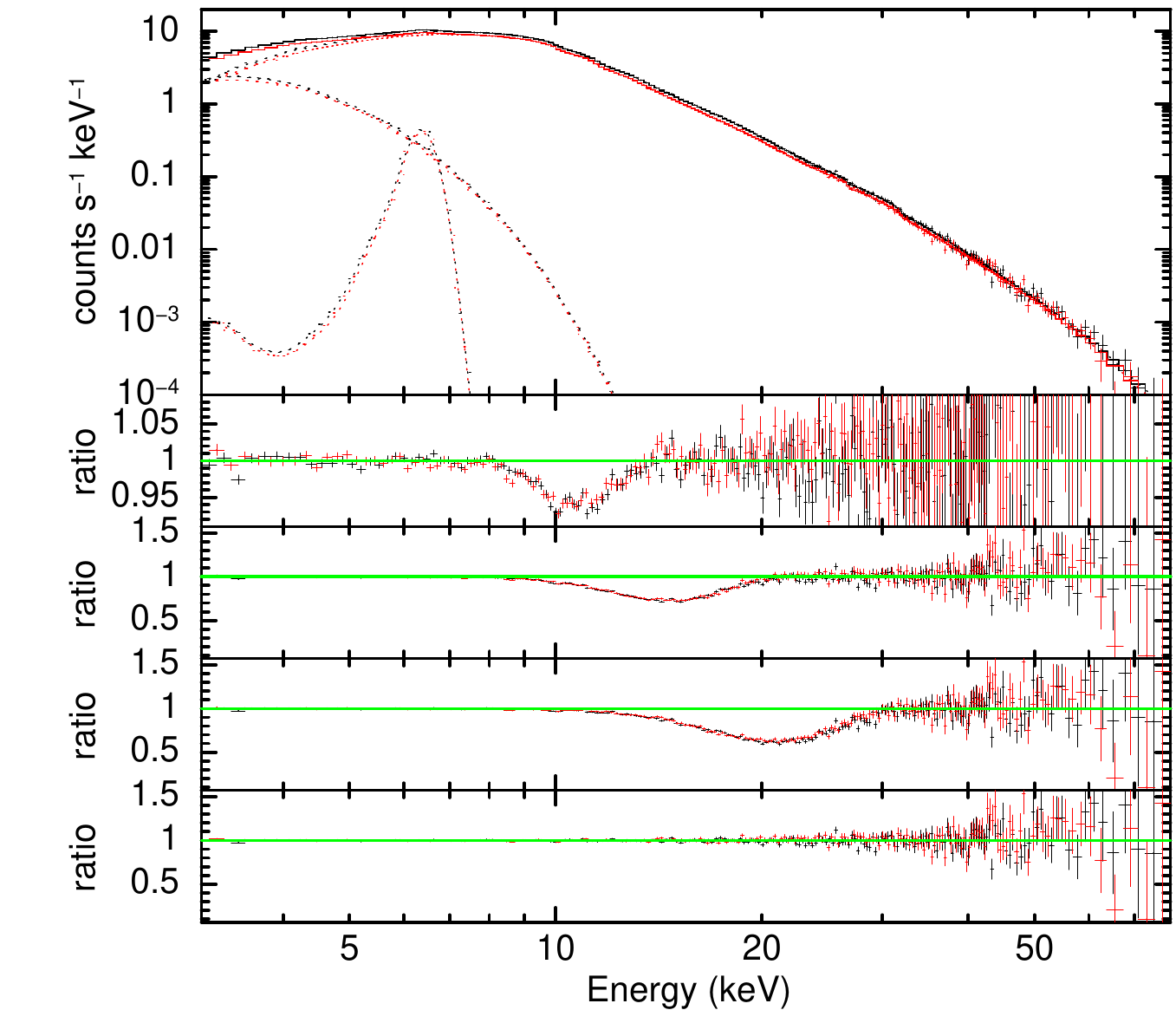}
    \caption{Spectrum and best fit spectral model for ObsID 04. The top panel consists of the best fit models along with the different components. Below this, the next three panels correspond to the residue from each of the three CRSFs at $\sim$12 keV, $\sim$16 keV, and $\sim$24 keV that were removed individually from the best fit model by setting the corresponding depth to zero. The bottom panel corresponds to the best fit model residue. The black points correspond to FPMA, while red points are for FPMB.
    \label{fig:Spectrum_2023_ID_4}}
\end{figure}

For the second observation ObsID 04 (Fig.~\ref{fig:Spectrum_2023_ID_4}), a similar spectral model was used. The broadband spectrum was modeled as \texttt{tbabs} $\times$ ( \texttt{compTT} + \texttt{bbodyrad} + \texttt{Gaussian} ) $\times$ multiple \texttt{gabs}. The spectral absorption lines at $\sim$11 keV and the first harmonic at $\sim$21 keV along with the line at $\sim$15 keV were present. The best fit spectral model gave a reduced $\chi^2$ of 1.28 for 369 degrees of freedom. The 3$-$50 keV luminosity of the source was calculated to be $2.83\pm0.01$ $\times$ $10^{37}$ erg s$^{-1}$. The best fit spectral parameters for both the observations are given in Table~\ref{Table:Spectral parameters} with errors quoted at 90\% confidence levels.

\begin{table*}[]
\caption{Best fit parameters of the 4U 0115$+$63 spectral fitting from both observations.}
\centering
\begin{tabular}{ccccc}
\hline
Spectral Model           & Spectral Parameter    & Units                  & Obs ID 2                                                                      & Obs ID 4                                                                     \\ \hline
TBabs                    & $N_H$                 & $10^{22}$ & $4.11_{-0.92}^{+0.80}$                                                        & $1.77_{-1.57}^{+1.32}$                                                       \\
compTT                   & $T_0$                 & keV                    & $2.14_{-0.08}^{+0.12}$                                                        & $2.56_{-0.04}^{+0.05}$                                                       \\
                         & kT                    & keV                    & $7.87_{-0.28}^{+1.75}$                                                        & $61.26_{-37.17}^{+36.15}$                                                    \\
                         & $\tau$                &                        & $2.59_{-0.68}^{+0.62}$                                                        & $0.01_{-0.01}^{+0.16}$                                                       \\
                         & Norm                  &                        & $0.28_{-0.04}^{+0.09}$                                                        & $6.96_{-0.08}^{+0.08}$$\times$ $10^{-3}$ \\
bbodyrad                 & kT                    & keV                    & $0.61_{-0.02}^{+0.02}$                                                        & $0.66_{-0.03}^{+0.04}$                                                       \\
                         & Radius                & km                     & $35.51_{-4.50}^{+4.12}$                                                                     & $13.41_{-2.79}^{+2.73}$                                                                    \\
Gaussian                 & $E_{\mathrm{center}}$ & keV                    & $6.44_{-0.03}^{+0.03}$                                                        & $6.39_{-0.03}^{+0.03}$                                                       \\
                         & $\sigma$              & keV                    & $0.18_{-0.05}^{+0.05}$                                                        & $0.19_{-0.07}^{+0.06}$                                                       \\
                         & Norm$^a$              &                        & $3.24_{-0.62}^{+0.78}$ $\times$ $10^{-3}$                                     & $1.13_{-0.26}^{+0.32}$$\times$ $10^{-3}$ \\
CRSF$_1$                 & $E_{\mathrm{center}}$ & keV                    & $11.95_{-0.12}^{+0.11}$                                                       & $10.78_{-0.17}^{+0.22}$                                                      \\
                         & $\sigma$              & keV                    & $1.80_{-0.16}^{+0.16}$                                                        & $1.21_{-0.20}^{+0.13}$                                                       \\
                         & Depth                 & keV                    & $0.70_{-0.22}^{+0.30}$                                                        & $0.20_{-0.07}^{+0.09}$                                                       \\
CRSF$_2$                 & $E_{\mathrm{center}}$ & keV                    & $16.00_{-0.94}^{+0.58}$                                                       & $14.73_{-0.21}^{+0.26}$                                                      \\
                         & $\sigma$              & keV                    & $3.02_{-0.76}^{+1.06}$                                                        & $2.59_{-0.26}^{+0.26}$                                                       \\
                         & Depth                 & keV                    & $1.25_{-0.60}^{+1.23}$                                                        & $2.09_{-0.54}^{+0.59}$                                                       \\
CRSF$_3$                 & $E_{\mathrm{center}}$ & keV                    & $23.83_{-0.22}^{+0.24}$                                                       & $21.06_{-0.39}^{+0.49}$                                                      \\
                         & $\sigma$              & keV                    & $3.07_{-0.46}^{+0.38}$                                                        & $4.19_{-0.43}^{+0.47}$                                                       \\
                         & Depth                 & keV                    & $2.82_{-1.32}^{+0.79}$                                                        & $5.01_{-0.81}^{+0.99}$                                                       \\
CRSF$_4$                 & $E_{\mathrm{center}}$ & keV                    & $34.02_{-0.44}^{+1.27}$                                                       & $-$                                                                             \\
                         & $\sigma$              & keV                    & $3.83_{-0.69}^{+1.17}$                                                        &  $-$                                                                            \\
                         & Depth                 & keV                    & $2.86_{-1.60}^{+1.96}$                                                        & $-$                                                                             \\
CRSF$_5$                 & $E_{\mathrm{center}}$ & keV                    & $47.36_{-0.71}^{+1.27}$                                                       &  $-$                                                                            \\
                         & $\sigma$              & keV                    & $7.59_{-2.82}^{+0.41}$                                                        &  $-$                                                                            \\
                         & Depth                 & keV                    & $8.00_{-4.67}^{+2.00}$                                                         & $-$                                                                             \\
Cross Normalization      &                       &                        & $0.994_{-0.001}^{+0.001}$                                                     & $1.001_{-0.001}^{+0.001}$                                                    \\
Reduced $\chi^2$ (d.o.f) &                       &                        & 1.29 (421)                                                                    & 1.28 (369)                                                                   \\
Flux                     & (3$-$50 keV)          & erg cm$^{-2}$ s$^{-1}$ & $2.06_{-0.01}^{+0.01}$$\times$ $10^{-8}$  & $4.83_{-0.01}^{+0.01}$$\times$ $10^{-9}$ \\
Luminosity               & (3$-$50 keV)          & erg s$^{-1}$           & $12.08_{-0.06}^{+0.02}$$\times$ $10^{37}$ & $2.83_{-0.01}^{+0.01}$$\times$ $10^{37}$ \\ \hline
\end{tabular}
\tablefoot{The errors are at a 90\% confidence interval. \tablefoottext{a}{The normalization factor is in units of total photon cm$^{-2}$ s$^{-1}$ in the line.}

}
\label{Table:Spectral parameters}
\end{table*}

\section{Results}

From the two \textit{NuSTAR} observations of 4U 0115$+$63 during the 2023 type II outburst, the spin period was measured to be 3.61456(1) s and 3.61309(1) s, respectively. The source luminosity changed by a factor of four between the two observations. There was a significant change in the pulse profile across the two observations. The pulse profile for both observations also showed variations with energy. The pulse profile were similar above 30 keV, while the profiles were substantially different below 30 keV. The pulse fraction for both observations showed an increasing trend with energy, with the pulse fraction reaching 80\% around 40 keV, similar to that observed during previous outbursts with the \textrm{RXTE} and \textrm{INTEGRAL} telescope~\citep{Tsygankov_PF_evolution_2007}. 

The spectrum of 4U 0115$+$63 was modeled with a Comptonized plasma model along with a soft thermal blackbody component with an iron emission line. There are five cyclotron lines present in the first observation and three in the second, the details of which are provided in Table~\ref{Table:Spectral parameters}. The two fundamental cyclotron lines were detected at 12 and 16 keV, along with their harmonics at 24, 34, and 48 keV. There was significant evolution of the CRSF line center energy with X-ray flux; we discuss this phenomena in detail in the discussion section. All the cyclotron lines in both observations were present in all the pulse phases of the NS. 

\section{Discussion}

The source 4U 0115$+$63 has undergone multiple type II outbursts since 1978, the most recent one being in 2023. The cyclotron line near $\sim$12 keV and the first harmonic near 20 keV were discovered during the 1978 outburst with the \textrm{HEAO 1} observatory~\citep{Wheaton_1979, White_1983} and were subsequently confirmed during the 1990 outburst with \textrm{GINGA} telescope~\citep{Mihara_1998}. The second, third, and forth harmonics of the 12 keV line near 34, 46, and 57 keV were detected during the 1999 outburst by both \textrm{RXTE} and \textit{BeppoSAX}~\citep{Heindl_1999, Santangelo_1999, Ferrigno_2009}. Although the fundamental energy of the cyclotron line provides the magnitude of the magnetic field strength, it is through the detection and analysis of the harmonic lines that we can obtain the geometrical configuration of the magnetic field~\citep{CRSF_theory_Harding_1991, CRSF_theory_Gochez_2000, CRSF_theory_Schonherr_2007}.

\subsection*{Variation in the cyclotron line energy with luminosity}

The exact luminosity dependence of cyclotron line features present in 4U 0115$+$63 has been a subject of much debate. The line center of the CRSFs changed from $\sim$11 keV at a high luminosity during the peak of the 1999 outburst to $\geq$15 keV at a lower luminosity~\citep{Nakajima_2006}. However, subsequent works have refuted this claim and attributed the result to inaccurate spectral modeling~\citep{Muller_2013, Boldin_2013}.

An updated spectral model was introduced by \citet{Suzaku_2008_outburst_Iyer_2015}, which simultaneously modeled the two fundamental cyclotron lines near 12 and 16 keV, along with the harmonics of these lines. This gave a good fit of the \textrm{RXTE}, \textit{Suzaku}, \textit{Swift,} and \textrm{INTEGRAL} data from the 2011 outburst. The two CRSFs at 12 and 16 keV cannot be harmonically related so they were considered as two fundamental CRSFs. Subsequent examinations of the source with \textit{NuSTAR} during the 2015~\citep{NuSTAR_2015_outburst_Liu_2020} and 2023 outbursts (present work) have affirmed the concurrent presence of the two fundamental lines and their harmonics in 4U 0115$+$63. The revised spectral model of 4U 0115$+$63 therefore requires a re-investigation of the previously reported variations in the centroid of the cyclotron line energy with the X-ray luminosity.

\begin{figure}
    \centering
    \includegraphics[width=0.45\textwidth]{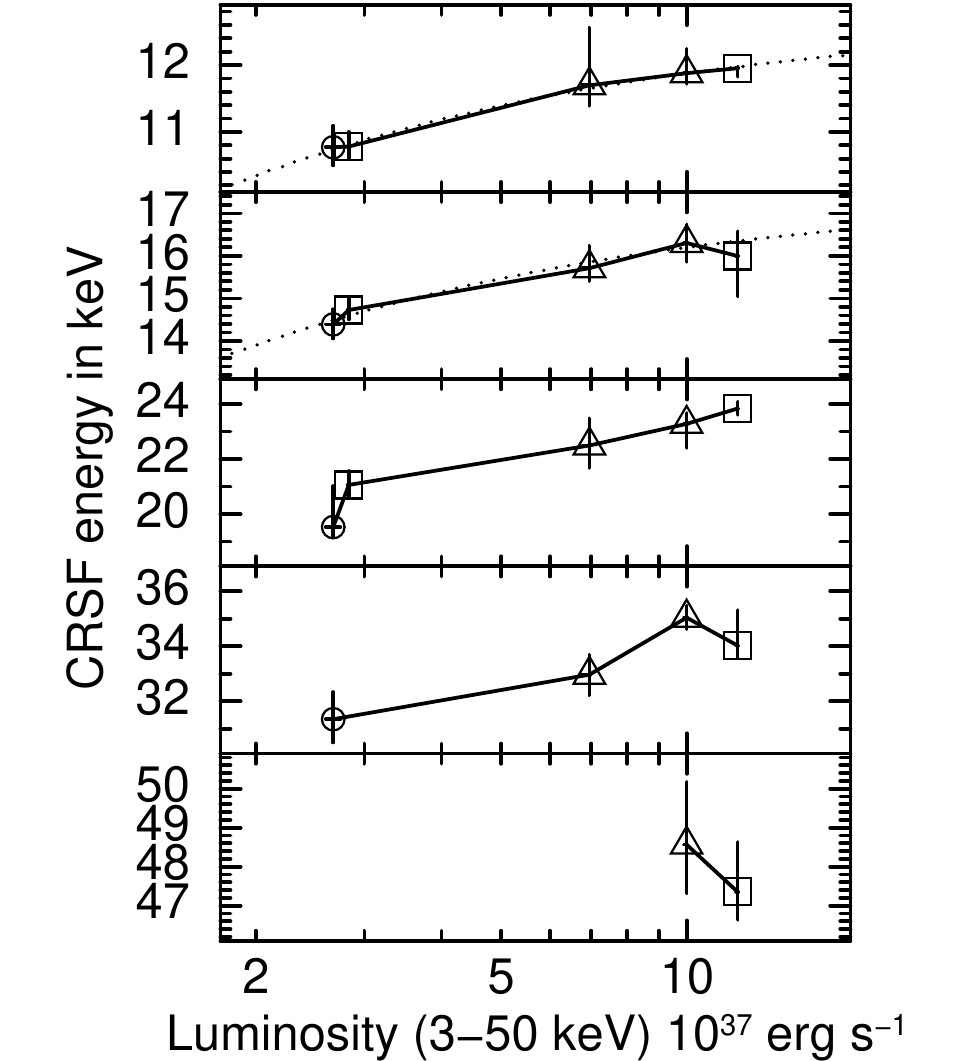}
    \caption{Variation in the line energies of different cyclotron lines with 3$-$50 keV luminosity of the source. Circles correspond to the 2011 outburst~\citep{Suzaku_2008_outburst_Iyer_2015}, triangles correspond to the 2015 outburst, and square points represent the results from the current work. The evolution of the 12, 16, 24, 34, and 48 keV are shown in different panels from top to bottom. The dotted lines in the top two panels show the predicted variation in the cyclotron line ($E_{cyc}$) and X-ray luminosity ($L_X$) according to a collisionless shock model.
    \label{fig:E_c_vs_lumin}}
\end{figure}

In Fig.~\ref{fig:E_c_vs_lumin} we have plotted the line center of multiple CRSFs in 4U 0115$+$63 as a function of the source luminosity in the 3$-$50 keV energy range. The circles correspond to the \textrm{RXTE} and \textit{Suzaku} data from the 2011 outburst (taken from \citeauthor{Suzaku_2008_outburst_Iyer_2015} \citeyear{Suzaku_2008_outburst_Iyer_2015}). The triangles correspond to the 2015 outburst and the squares represent the results from the 2023 outburst with \textit{NuSTAR}. The CRSF energy and the 3$-$50 keV luminosity values of \textit{NuSTAR} observations during the 2015 and 2023 outbursts were recalculated for this work. For this work, we considered the observations where the 12 and 16 keV lines were simultaneously considered and studied the variation in the line center with luminosity. As we can clearly see in Fig.~\ref{fig:E_c_vs_lumin}, there is a simple positive correlation between the line center of CRSFs and the source luminosity. While all the lines follow this behavior, the correlation is specially prominent for the two fundamental cyclotron lines near 12 and 16 keV. The increase in the energy of the line center of the two fundamental cyclotron lines with luminosity is about 10 percent. The change in CRSF with energy is found until $10^{38}$ erg/s for the 16 keV line and its harmonics as these CRSFs show a tentative negative correlation with luminosity beyond $10^{38}$ erg/s. 

The positive correlation between the cyclotron line central energy (\textit{E$_{cyc}$}) and luminosity (\textit{L$_{x}$}) was discovered in the persistent HMXB Her X-1~\citep{HerX1_CRSF_Staubert_2007}. Subsequently, an increase in \textit{E$_{cyc}$} with \textit{L$_{x}$} has been found in Vela X-1~\citep{VelaX1_CRSF_Furst_2014, VelaX1_CRSF_LaParola_2016}, A 0535+63~\citep{Muller_A0535p26_pulse2pulse_CRSF_2013, Caballero_A0535p26_flare_CRSF_2008}, GX 304$-$1~\citep{GX304m1_CRSF_Malacaria_2015, GX304m1_CRSF_Rothschild_2017}, Cep X$-$4~\citep{CepX4_CRSF_Vybornov_2017}, and Swift J1626.6$-$5156~\citep{Swift_J16266m5156_CRSF_DeCesar_2013}. Her X-1 and Vela X-1 are the two persistent HMXB sources that show a positive correlation in \textit{E$_{cyc}$} with \textit{L$_{x}$} in the luminosity range of $\sim$ 0.1 to 3 $\times$ $10^{37}$ erg s$^{-1}$. The positive correlation in Cep X-4 was observed in a pulse-to-pulse analysis~\citep{CepX4_CRSF_Vybornov_2017}, which was not seen in the phase average results across the multiple outbursts of the source~\citep{CepX4_CRSF_KallolM_2021}. For A 0535+63, \textit{E$_{cyc}$} was found to be more in a flare observed before the peak of the 2005 outburst compared to the line energy during the main outburst~\citep{Caballero_A0535p26_flare_CRSF_2008}. A positive correlation was observed in the pulse-to-pulse analysis of the 2010 outburst of the source~\citep{Muller_A0535p26_pulse2pulse_CRSF_2013}; however, recent observations of the source during its 2020 outburst show a negative correlation beyond a critical luminosity~\citep{A0535p26_Kong_2021}. 4U 0115$+$63 is the only HMXB that shows a positive correlation between cyclotron line and luminosity over a very wide range of X-ray luminosities from $\sim$ 2.6 to 12 $\times$ $10^{37}$ erg s$^{-1}$.

A positive correlation is observed in cyclotron line energy with the luminosity in the collisionless shock regime~\citep{CRSF_Becker_2012}, and beyond a critical luminosity the relation changes to a negative correlation~\citep{CRSF_review_Staubert_2019}. The variation in the line center with flux is entirely due to the magnetic field strength at the height above the NS surface responsible for the production of the CRSFs. The scaling of the magnetic field strength with flux could well explain the luminosity dependence in Her X-1~\citep{HerX1_CRSF_Staubert_2007}. In sources displaying a positive correlation, there is an increase in the height of the region in the accretion mound near the poles of the NS where the CRSF is produced with a decrease in luminosity. While at a higher height in the accretion mount, the magnetic field is smaller leading to a lesser value for the CRSF centroid energy.

The centroid energy of the CRSF is directly proportional to the magnetic field strength at the line-forming region (eq.~\ref{E_cyc_vs_B_12}). Assuming a dipolar magnetic field, the magnetic field strength ($B$) is inversely proportional to the cube of distance ($r$) from the center of the NS. Therefore, the energy of the cyclotron line as a function of the height from the NS surface can be written as

\begin{equation}
    E_{cyc}(H_S) = E_0 \left( \frac{R_{NS}}{ H_S + R_{NS}} \right)^3 \times \frac{1+z(R_{NS})}{1+z(H_S)}
    \label{eq:CRSF_vs_H_s}
,\end{equation}

where $R_{NS}$ is the radius of the NS, $E_0$ is the value of the CRSF energy on the NS surface, $z(R_{NS})$ is the gravitational redshift on the NS surface, and $z(H_S)$ is the gravitational redshift at a height $H_S$ from the NS surface. The value of gravitational redshift $z(H_S)$ at a distance $H_S$ from the surface of the NS of radius $R_{NS}$ is written as

\begin{equation}
    1 + z(H_S) = {\left(1 - \frac{2GM}{\left(R_{NS} + H_S \right)c^2}\right)^{-1/2}}
    \label{eq:z_H_S}
,\end{equation}

where G is the gravitational constant and c is the speed of light in vacuum. The height of the shock-forming region can be written as a function of the X-ray luminosity ($L_X$) as~\citep{CepX4_CRSF_Vybornov_2017} 

\begin{equation}
    \frac{H_s (L_{X})}{R_{NS}} = K L_X^{-\alpha}
    \label{eq:H_s_vs_L_x}
,\end{equation}

where $\alpha$ is the power law index and $K$ is the proportionality constant. For quasi-spherical settling accretion, $\alpha$ $=$ $9/11$ and $\alpha$ $=$ $5/7$ for quasi-spherical Bondi accretion or disk accretion~\citep{Shakura_Accretion_NS_2012, CepX4_CRSF_Vybornov_2017}. Combining eq.~\ref{eq:CRSF_vs_H_s}, eq~\ref{eq:z_H_S} and eq.~\ref{eq:H_s_vs_L_x}, the variation in the cyclotron line energy with X-ray luminosity for the collisionless shock model can be written as

\begin{equation}
    E_{cyc}(L_X) = \frac{E_0}{\left(  1 + K L_X^{-\alpha} \right)^{3}} \times \frac{\left(1 - \frac{2GM}{R_{NS}c^2}\right)^{-1/2}}{\left(1 - \frac{2GM}{R_{NS}\left(1 + K L_X^{-\alpha}\right)c^2}\right)^{-1/2}}
    \label{eq:CRSF_vs_Lx}
,\end{equation}

where G is the gravitational constant, c is the speed of light in vacuum, $K$ and $E_0$ are the free parameters, and $\alpha$ is taken as $5/7$ corresponding to disk or Bondi quasi-spherical accretion. The best fit values of the two free parameters $E_0$ and K were calculated to be 12.7 $\pm$ 0.1. keV and 0.130 $\pm$ 0.004 (erg/s)$^\alpha$  for the 12 keV line, respectively. The magnetic field strength corresponding to the surface value of the CRSF energy was calculated from eq.~\ref{E_cyc_vs_B_12} to be 1.4 $\times$ $10^{12}$ G. For the 16 keV line, the best fit value was calculated to be 17.4 $\pm$ 0.3. keV and 0.145 $\pm$ 0.017 (erg/s)$^\alpha$ for $\alpha$ and $K$, respectively. The fit was performed neglecting the highest luminosity data point for the 16 keV line, as there is a hint of negative correlation between $E_{cyc}$ and $L_X$ beyond $10^{38}$ erg/s. The magnetic field strength corresponding to the surface value of the 16 keV CRSF energy was calculated to be 2.0 $\times$ $10^{12}$ G. The variation in the cyclotron line ($E_{cyc}$) with X-ray luminosity ($L_X$) for the 12 (red) and 16 (black) keV CRSF are shown in the top two panels of Fig.~\ref{fig:E_c_vs_lumin}. The collisionless shock model has been successful in describing the luminosity variation in the cyclotron line for sources such as GX 304$-$1~\citep{GX304m1_CRSF_Rothschild_2017} and Cep X$-$4~\citep{CepX4_CRSF_Vybornov_2017}.

The correlation between $E_{cyc}$ and $L_X$ shows a possible change from positive to negative beyond $10^{38}$ erg/s for the fundamental CRSFs at 16 keV. A similar behavior can also be observed in the first (34 keV) and second (48 keV) harmonic of the same fundamental cyclotron line. However, with only one data point, this trend cannot be established beyond a reasonable doubt and more high luminosity observations of 4U 0115$+$63 are needed to firmly establish the negative correlation at a high luminosity. A negative correlation between $E_{cyc}$ and $L_X$ is observed in the radiative shock regime~\citep{CRSF_Becker_2012} and the transition from positive correlation occurs beyond a critical luminosity. The value of critical luminosity for 4U 0115$+$63 is $\sim$ $10^{38}$ erg/s. The change from a positive to negative correlation beyond a luminosity is not seen in the 12 keV fundamental CRSFs as well as its first harmonic at 24 keV. A possible transition from a positive to negative correlation of $E_{cyc}$ and $L_X$ has been observed in XTE J1946$+$274 for one of the magnetic poles~\citep{Ashwin_J1946_HXMT_2024}. A similar change in correlation was observed in A 0535+262~\citep{A0535p26_Kong_2021} during its 2020 outburst and also in V 0332$+$53~\citep{V0332p53_Vybornov_2018} from a pulse amplitude resolved analysis. The correlation was observed in both sources only after correcting for the time variation in the CRSF line center.

Two fundamental CRSFs originate from two line-forming regions in the atmosphere of the NS. The two fundamental lines can originate from the two line-forming regions at different heights along the same pole of the NS, as has been observed in GX 301$-$2~\citep{GX301m2_CRSF_Furst_2018}. The alternate model for the origin of the two fundamental CRSFs is that they are formed in two different magnetic poles of the NS. This Two Pole Cyclotron line Model (TPCM) has been preferred from the results of the 2008~\citep{Suzaku_2008_outburst_Iyer_2015} and 2015 outburst~\citep{NuSTAR_2015_outburst_Liu_2020}. The TPCM origin of the two fundamental lines can be due to the non-dipolar structure of the magnetic field, misalignment of the magnetic poles, different accretion rates for the two magnetic poles, or a different structure for the accretion column at the two magnetic poles. The increase in the CRSF energy of the two fundamental lines are consistent with the collisionless shock model, and in that scenario, the two CRSFs probably originate from aligned magnetic poles with a different accretion column geometry, accretion rate, or from misaligned magnetic poles of the NS. We also tried to study the origin of the two fundamental lines using spin-phase resolved spectroscopy. Even though the two fundamental cyclotron lines and their harmonics were detected in all the phase bins, no concrete conclusion could be drawn from the data regarding the origin of the two fundamental CRSFs.

\section{Summary}

In this work, we report the results from an analysis of the two \textit{NuSTAR} observations of 4U 0115$+$63 during the 2023 outburst of the source. Two fundamental cyclotron lines near 12 and 16 keV were simultaneously observed in the source similar to the 2011~\citep{Suzaku_2008_outburst_Iyer_2015} and 2015 outburst~\citep{NuSTAR_2015_outburst_Liu_2020}. Multiple harmonics of the two fundamental lines were present near 24, 34, and 47 keV.

The CRSFs present in 4U 0115$+$63 show a positive correlation with luminosity. There is a hint of a negative correlation beyond the luminosity value of $10^{38}$ erg/s for the 16 keV CRSF as well as its harmonics. No such behavior was observed for the fundamental and first harmonic of the 12 keV CRSF. The collisionless shock model could well describe the luminosity variation in the cyclotron line energy. Assuming that the two fundamental CRSFs are produced at the two poles of the NS~\citep{NuSTAR_2015_outburst_Liu_2020}, the dipolar magnetic field strength of the two poles are calculated to be 1.4 and 2.0 $\times$ $10^{12}$ G. We also comment on the different scenarios as to the  origin of the two fundamental CRSFs. However, future observations across a wide range of luminosities is essential to distinguish between the possible scenarios.

\begin{acknowledgements}
      We thank the referee for the useful comments that improved the quality of this paper. KR would like to thank Rahul Sharma and Ashwin Devaraj for their inputs during the preparing of the manuscript for this project. This research made use of data obtained with \textit{NuSTAR}, a project led by Caltech, funded by NASA and managed by NASA/JPL.
\end{acknowledgements}

\bibliographystyle{aa}
\bibliography{bibliography}

\end{document}